# Bandgap measurement of high refractive index materials by off-axis EELS

Maryam Vatanparast[1], Ricardo Egoavil[2], Turid W. Reenaas[1], Johan Verbeeck[2], Randi Holmestad[1] and Per Erik Vullum[1, 3]

1. *Department of Physics, Norwegian University of Science and Technology, NO-7491 Trondheim, Norway*
2. *EMAT, University of Antwerp, Groenenborgerlaan 171, 2020 Antwerp, Belgium*
3. *SINTEF Materials and Chemistry, 7065 Trondheim, Norway*

**Abstract**

In the present work, Cs aberration corrected and monochromated scanning transmission electron microscopy electron energy loss spectroscopy STEM-EELS has been used to explore experimental set-ups that allows bandgaps of high refractive index materials to be determined. Semi-convergence and –collection angles in the µrad range were combined with off-axis or dark field EELS to avoid relativistic losses and guided light modes in the low loss range to contribute to the acquired EEL spectra. Off-axis EELS further suppressed the zero loss peak and the tail of the zero loss peak. The bandgap of several GaAs-based materials were successfully determined by direct inspection and without any background subtraction of the EEL spectra. The presented set-up does not require that the acceleration voltage is set to below the Cerenkov limit and can be applied over the entire acceleration voltage range of modern TEMs and for a wide range of specimen thicknesses.

# 1. Introduction

The developments of aberration corrected and monochromated TEMs were expected to revolutionize the access to information about semiconductor bandgaps and optical properties at the nanoscale. Technically, this is the case, scanning transmission electron microscopy (STEM) with a Cs aberration corrected and monochromated electron beam can in combination with electron energy loss spectroscopy (EELS) be used to collect spectroscopy maps that have atomic spatial resolution and an energy resolution that is in the tens of meV range. Hence, the various bandgaps of nanoscale structures can in principle be mapped both easily and quickly. Moreover, since the low-loss region of the EEL spectrum is described classically by the energy-loss function that is determined by the material's dielectric function [1-3], Kramer-Kronig analysis can be used to determine the material's optical response.

However, many semiconductors are also high refractive index materials. The speed of the electrons in the sample material scales with the refractive index. This means that the charge carrier velocity might exceed the phase velocity of light in the material. As a consequence, Cerenkov losses (CLs) or Cerenkov photons are generated in the low energy range of the loss spectrum [4-5]. Besides the pure relativistic CLs, further low-energy losses appear due to retardation of the beam electrons from surface and interface plasmons (these losses also have relativistic contributions) and from excitation of guided light modes [6]. The probability of generating Cerenkov radiation increases with acceleration voltage (i.e. the speed of the electrons), but disappears below a critical acceleration voltage (Cerenkov limit) determined by the sample's maximum value of the real part of the dielectric function [6]. Surface and interface plasmon losses and excitation of guided light modes will still be present at acceleration voltages below the Cerenkov limit [6]. For a large and important group of high refractive index materials, such as Ge, Si, GaAs, GaP and many other semiconductors, the Cerenkov limit is below the typical acceleration voltage range of modern TEMs. The Cerenkov limit for GaAs is only 11 keV if the maximum value of the refractive index is used, and 20.6 keV if the refractive index at the bandgap is used [7].

Another complication is the extremely low signal of the interband transitions compared to the intensity of the zero loss peak (ZLP). Even if the full width half maximum (FWHM) of the ZLP is very low, a careful procedure for handling the tail of the ZLP is important in order to extract a correct bandgap signal [8]. A vacuum-recorded ZLP is always different from the ZLP in an EEL spectrum acquired from the sample due to phonon and exciton losses, as well as imperfections in the spectrometer, converting elastic scattering which changing the incoming angles into seemingly inelastic signals. Generally, a fitting procedure for the ZLP, and especially the tail of the ZLP, is preferred. Several methods have been attempted to extract the bandgap by various fitting routines. These routines include the use of a Lorentzian function [9], a seven parameter fit function for ZLP deconvolution [10], a power law background in front of the bandgap [3], and mirroring the left-side tail of the ZLP to the energy loss side [11]. Even if some of these fitting methods can remove phonon and exciton losses in the ZLP tail, relativistic losses present both below and above the bandgap of high refractive index materials cannot reliably be removed by any fitting routine [12].

In normal STEM, the semi-convergence angles of the beam are normally so large [~several mrad] that the convergent beam electron diffraction (CBED) discs have a significant overlap. This also means that CLs, surface losses and guided light modes are present in the entire diffraction plane. Hence, normal STEM cannot be used to measure bandgaps in high refractive index materials in the standard acceleration voltage range of 60 – 300 kV for most modern TEMs. However, in this work we use a monochromated and probe corrected beam in low-magnification (Low-Mag) STEM mode to directly map the bandgap of various GaAs-based multilayer structures. We utilize that CLs and light guided modes are only present at scattering angles below a few tens of µrad [12-15].

The bandgap of GaAs has been retrieved in the literature before. In recent work, it has been argued that unwanted spectral contributions can be avoided by working below the Cerenkov limit [1], and that this simplifies credible measurement of bandgap energies. Nevertheless, in the relativistic simulations of valence electron energy-loss spectra of GaAs done by Rolf Erni it was found that the spectra could still be affected by excitation of guided light modes and retardation effects even with electron energies below the Cerenkov limit [13]. These energy losses are represented by a smooth thickness-dependent background, which is difficult to treat when removing the zero-loss peak from the spectra. Attempts to remove this background might be critical for materials with a weak bandgap feature like GaAs and could affect the measured bandgap energy.

In low magnification STEM, semi-convergence and –collection angles in the µrad range are used. Off-axis conditions, i.e. dark field EELS, is further used to suppress the ZLP and to be outside the angular range where CLs and light guided modes hit the spectrometer. Furthermore, the low semi-convergence and -collection angles suppress the detection of phonons (they have a large angular distribution) and interband transitions with a significant momentum transfer. The presented dark field set-up gives EEL spectra where the bandgap signal is extracted directly from the raw spectra without any ZLP background subtraction procedure due to the suppression of the ZLP and its tail. Dark field EELS means that we detect electrons with a momentum transfer. However, because the semi-angles are far down into the µrad range, we can still detect a part of reciprocal space that is close to the centre of the first Brillouin zone (BZ), the Gamma point (Γ), and far from any of the BZ boundaries. Since the valence and conduction bands both usually are very flat at the BZ center, the measured bandgap is still equal or very close to the material's direct bandgap. One of the benefits of the presented methodology is that bandgap of high refractive index materials can be determined at any acceleration voltage and over the entire range of conventional TEM specimen thicknesses.

After presenting the different experimental set-ups, we will in this paper systematically compare the results from the different methods of band gap measurements of GaAs-based materials.

## 2. Materials and method

The GaAs-based materials characterized in this work were grown by molecular beam epitaxy (MBE). A high angle annular dark field (HAADF) STEM image of the sample structure is shown in Fig. 1 a). A 400 nm thick $Al_{0.25}Ga_{0.75}As$ layer is grown on top of a (001) GaAs substrate followed by a stack of 20 quantum dot (QD) layers. Each of the QD layers were grown as 2 monolayers thick InAs and they are separated by 20 nm thick $Al_{0.25}Ga_{0.75}N_xAs_{1-x}$ (x < 0.01) spacer layers. On top of the QD stack there is a 330 nm thick $Al_{0.25}Ga_{0.75}As$ layer followed by a 50 nm thick GaAs top layer.

Cross-section TEM samples were prepared both by focused ion beam (FIB) and by Ar ion-milling. The FIB samples were prepared by a FEI Helios Nanolab dual-beam FIB equipped with an Oxford Omniprobe. The coarse $Ga^+$ ion-beam thinning was done at 30 kV acceleration voltage followed by final thinning at 5 kV and 2 kV. A high resolution HAADF STEM image from one of the InAs QD layers is shown in Fig. 1 b) and demonstrates that the samples made purely by FIB have very minor beam damage from the sample preparation. A PIPS II was used to thin the $Ar^+$ ion-milled samples. Coarse thinning was performed at 3 kV acceleration voltage before progressively reducing the acceleration voltage, finishing at 100 eV. The samples were cooled by liquid $N_2$ during milling.

The HAADF STEM images in Fig. 1 were taken with a double Cs corrected, coldFEG JEOL ARM 200CF, operated at 200 kV. EELS was performed with a double Cs corrected, monochromated Titan cube, operated at 80 or 120 kV and equipped with a Gatan Imaging Filter (GIF) Quantum ERS spectrometer. An energy dispersion of 0.01 eV/pixel, and a 2.5 mm GIF entrance aperture were used at all times. The GaAs-based material was oriented a few degrees away from the [110] zone axis to avoide strong channeling effects, and to make sure that we are off most Kikuchi bands in the off-axis set-up.

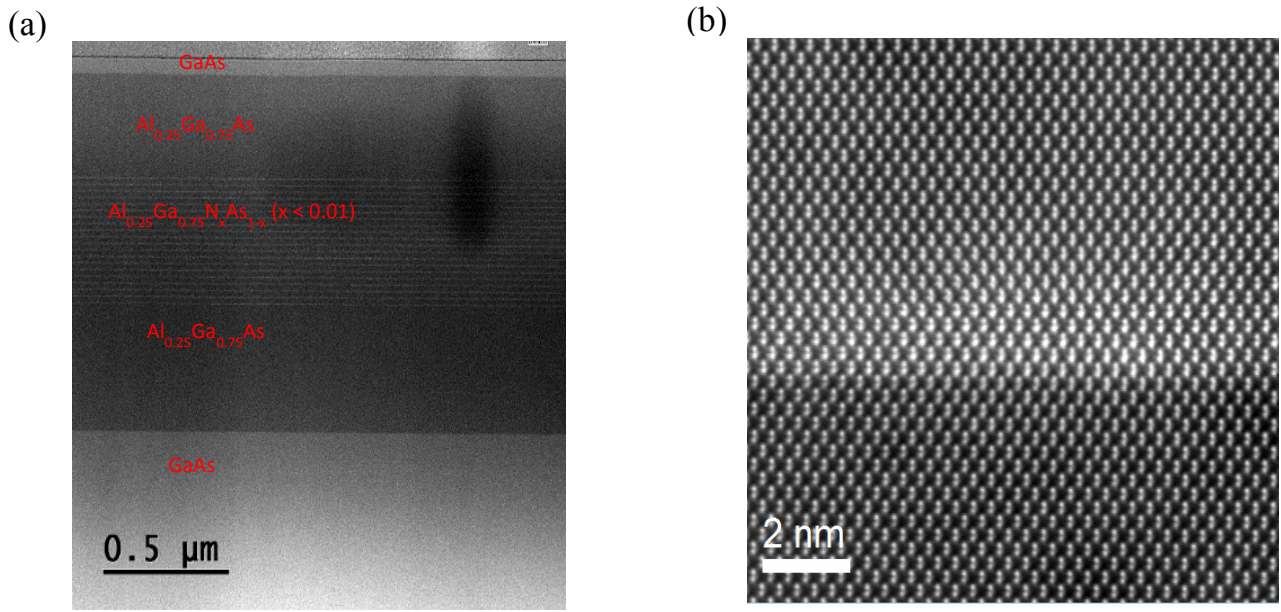

**Fig. 1. (a)** A high angle annular dark field (HAADF) STEM image of the sample structure. A 400 nm thick $Al_{0.25}Ga_{0.75}As$ layer is grown on top of a (001) GaAs substrate followed by a stack of 20 quantum dot (QD) layers. Each of the QD layers were grown as 2 monolayers thick InAs and they are separated by 20 nm thick $Al_{0.25}Ga_{0.75}N_xAs_{1-x}$ (x < 0.01) spacer layers. On top of the QD stack there is a 330 nm thick $Al_{0.25}Ga_{0.75}As$ layer followed by a 50 nm thick GaAs top layer. (b) High magnification HAADF-STEM image of an InAs quantum dot from layer 1, In is heavier than Ga, giving higher contrast.

Several set-ups, combining both on-axis and off-axis, normal and low magnification STEM mode with the beam on and closely outside (Aloof) the sample, were conducted to find a set-up that would produce credible bandgap measurements.

In normal STEM mode, the semi-convergence and semi-collection angles are both in the milliradian (mrad) range. The simulated convergent beam electron diffraction (CBED) pattern in Fig. 2 a) shows the size of and the distance between the diffraction discs for GaAs along the [110] zone axis, and for a semi-convergence angle of 12.6 mrad at 80 kV. This semi-convergence angle combined with a 7.7 mrad semi-collection angle (2.5 mm GIF entrance aperture and 2100 mm camera length at 80 kV) give EEL spectra that probe the entire first BZ, in addition to significant parts of second BZ. However, electrons that have received higher momentum transfer will also go onto the spectrometer due the size of the CBED discs. In the Low-Mag STEM set-up the electron beam is much more parallel than in the normal STEM set-up and allows for semi-convergence and semi- collection angles that both are far down into the μrad range. Simulations of the size of and the distance between the diffraction discs for GaAs along the [110] zone axis at 80 kV in the Low-Mag mode are shown in Fig. 2 b). The semi-convergence angle in these simulations is 0.20 mrad and identical to the angle used in our experiments (defined by the 50 μm in diameter condenser aperture). In Fig. 3, the 0.20 mrad semi-convergence angle is combined with our experimental semi-collection angle of 0.05 mrad (2.5 mm GIF entrance aperture combined with a 6100 mm camera length). These semi-angles are now so small that we practically get signal only from the center of the first Brillouin zone, assuming on-axis conditions. In order to reach off-axis conditions or dark field EELS conditions where the entire 000 diffraction disc falls outside the spectrometer, the diffraction pattern needs to be shifted at least 0.25 mrad (See Fig. 3). However, under such off-axis conditions the region of reciprocal space detected by the spectrometer is still very close to the center of the 1$^{st}$ BZ. For GaAs along the [110] zone axis, the region of reciprocal space that goes onto the spectrometer is less than 5% away from the Γ point (see Figs. 2 and 3 b)). Since the valence and conduction bands often have parabolic shape at the zone center, our off-axis conditions should still detect a signal that is very close to the direct bandgap. The relativistic losses, surface modes and bulk waveguide modes all are extremely forward scattered, and exist inside a narrow solid angle that extends out to a few tens of mrads. As such, our off-axis conditions are close enough to the center of the first Brillouin zone to detect bandgap excitations of almost direct bandgap transitions, but still outside the narrow angular range where the unwanted signals from relativistic losses etc. compromise the EEL spectra.

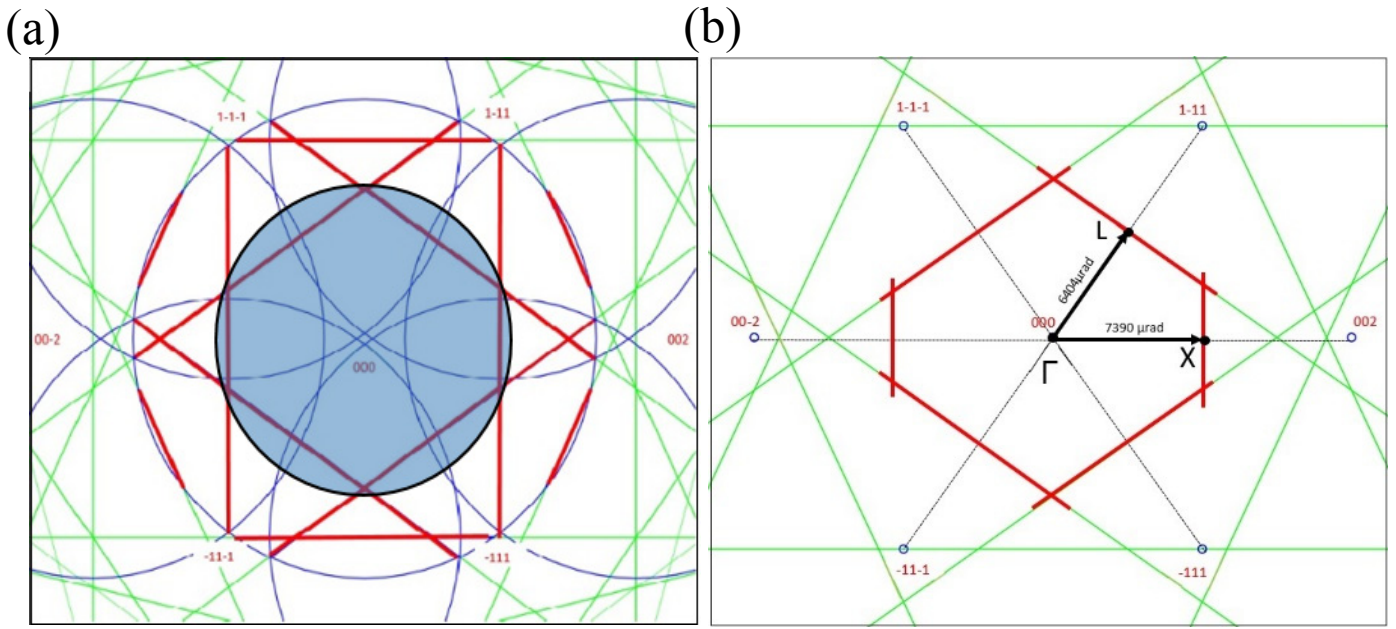

**Fig. 2.** Simulated CBED patterns from the [011] zone axis orientation in GaAs. (a) and (b) show the size of and the distance between the diffraction discs for GaAs in normal and Low-Mag STEM set-ups, respectively. In the normal STEM mode, a semi-convergence angle of 12.6 mrad combined with a 7.7 mrad semi-collection angle give EEL spectra that probe the entire first BZ, in addition to significant parts of second BZ. In the Low-Mag STEM set-up, the semi-convergence and -collection angles are both far down into the µrad range. The 0.20 mrad semi-convergence angle is combined with 0.05 semi-collection angle, and we practically get signal from the center of the first BZ. Green lines are Kikuchi lines, blue circles show the size of CBED disks and the BZ boundaries are shown with red lines. The blue shaded circle shows the semi- collection angle in the normal STEM set-up [20].

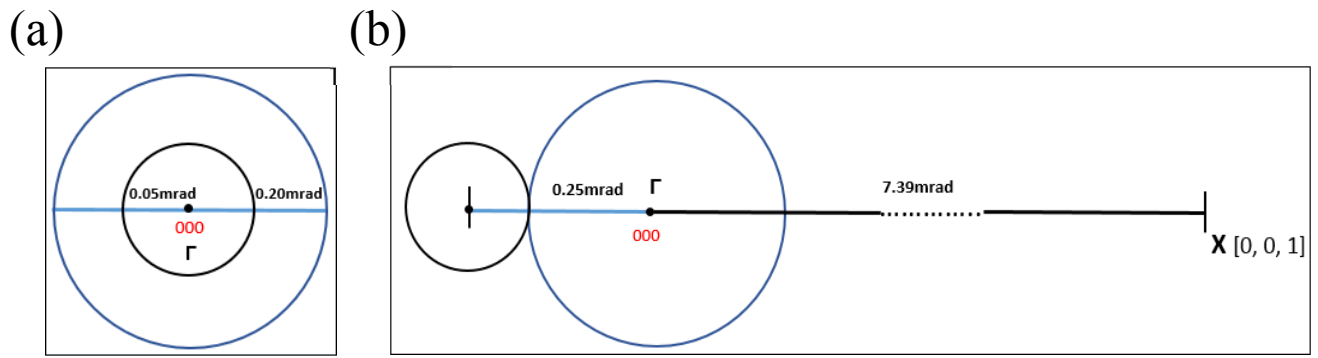

**Fig. 3.** The 0.20 mrad semi-convergence angle with a 50 µm CL3 aperture (blue circle) combined with 0.05 semi-collection angle (black circle) in Low-Mag on-axis (a) and Low-Mag off-axis (b) set-ups are illustrated. In off-axis they are scaled according to the 1$^{st}$ Brillouin zone boundary, X point, along the [0, 0, 1] direction. In off-axis or dark field EELS conditions the entire 000 diffraction disc falls outside the spectrometer, when the diffraction pattern is shifted at least 0.25 mrad. Under such off-axis conditions the region of the reciprocal space detected by the spectrometer is still very close to the center of the 1$^{st}$ BZ. For GaAs along the [110] zone axis, the region of reciprocal space that goes onto the spectrometer is less than 5% away from the Gamma point.

Aloof measurements [16] were performed by positioning the beam in vacuum just outside the sample. The idea here was to exploit the delocalized nature of the bandgap and simultaneously supress the unwanted signal in the bandgap region that masks the signal from bandgap excitations. The possible advantage of aloof is that the tail of the ZLP and relativistic losses can be supressed. To create ideal TEM samples for aloof measurements the final TEM sample was mechanically broken by the tungsten Omniprobe needle in the FIB. This procedure was used to create a TEM sample with a sample facet parallel to the electron beam in order to maximize the interaction between the electron beam and the long-range part of the wave function that describes the sample. Furthermore, this procedure avoids any sample damage caused by the $Ar^+$ or the $Ga^+$ ion-beams. In the following result part, EEL spectra from the different set-ups are presented and analysed in order to find the best method for measuring bandgaps in high refractive index materials. Table 1 shows an overview of the different angles used in the different set-ups.

**Table 1.** Semi-convergence and collection angles used in the different set-ups tested in this work.

|  | Acceleration voltage (kV) | Semi-convergence angle (mrad) | Semi-collection angle (mrad) |
|---|---|---|---|
| **Normal STEM set-up (Fig. 1)** | 200 | 27 | - |
| **Normal STEM set-up (Fig. 5)** | 120 | 12.6 | 64.7 |
| **Normal STEM set-up (Figs. 2a, 4a, 4b and 8)** | 80 | 12.6 | 7.7 |
| **Low-Mag STEM set-up (Fig. 6)** | 80 | 0.12 | 0.05 |
| **Low-Mag STEM set-up (Figs. 2b and 7)** | 80 | 0.20 | 0.05 |

# 3. Results and discussion

### 3.1. Normal STEM on-axis

We first tried to duplicate earlier obtained results for the bandgap in GaAs [1,6] by using normal STEM mode. Measurements were performed at 60, 80 and 120 kV and with sample thicknesses in the range of 20 – 150 nm. Furthermore, dual EELS mode was used to maximize the signal to noise in the energy loss region of the bandgap, as the intense part of the ZLP was positioned outside the spectrometer in the "high loss" spectrum. Independent of TEM sample thickness and acceleration voltage, no distinct feature due to bandgap excitations can directly be observed in any of the raw spectra. One such spectrum, acquired at 80 kV is shown in Fig. 4 a). Several background subtraction methods were approached in order to reliably subtract the tail of the ZLP. However, methods such as mirroring the left side (negative energy losses) of the zero loss to the right side (positive energy losses), or to use the ZLP acquired in vacuum to subtract the background, both fail. The reason these methods fail can clearly be seen in Fig. 5 where an EEL spectrum acquired in vacuum is compared with the spectrum acquired under the same conditions on the GaAs-based structure. It is not only the tail of the ZLP in the loss region before the bandgap energy that changes significantly, but even the full width at half maximum (FWHM) of the ZLP changes from 0.10 eV in vacuum to 0.14 eV on the sample. A large semi-collection angle of 64.7 mrad was used during the acquisition of the spectra in Fig. 5. Hence, phonon and multiphonon excitations will contribute strongly to the spectrum, in addition to other significant losses such as intraband transitions, relativistic losses, guided light modes and surface and interface related losses in the pre bandgap region. Furthermore, the relative strength of most of these losses, which mask the signal from bandgap excitations, is dependent on the sample's chemical composition, sample thickness and crystal orientation as well as acceleration voltage and semi-convergence and –collection angles.

The EEL signals in Fig. 4 b) were produced with a standard "power law" background subtraction model. A power law function gives a very good fit to the spectrum in the entire range from ca. 1.0 – 2.5 eV loss if the width of the energy window chosen to fit the power law function is chosen wisely. However, such an approach can reproduce a bandgap that can take any value in the range 1.0 – 2.5, including the known direct bandgap of GaAs at 1.42 eV.

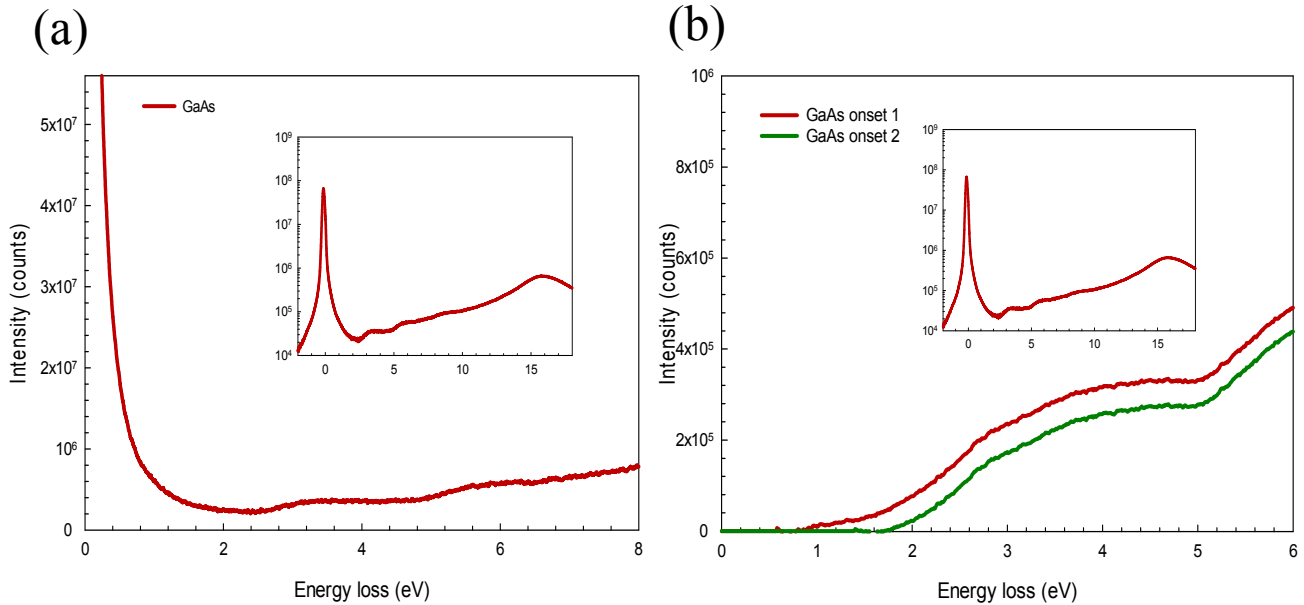

**Fig. 4.** (a) Raw EELS spectrum acquired from the normal STEM set-up at 80kV. No distinct feature due to bandgap excitations can be directly observed in any of the raw spectra. (b) Different bandgap onsets for GaAs in the normal STEM set-up after background subtraction. The EEL signals produced with a standard "power law" background subtraction model. A power law function gives a very good fit to the spectrum in the entire range from ca. 1.0 – 2.5 eV loss if the width of the energy window chosen to fit the power law function is chosen wisely. This approach can reproduce a bandgap that can take any value in the range between 1.0 – 2.5 eV, including the known direct bandgap of GaAs at 1.42 eV. The whole EELS spectrum is shown in logarithm scale in the inset.

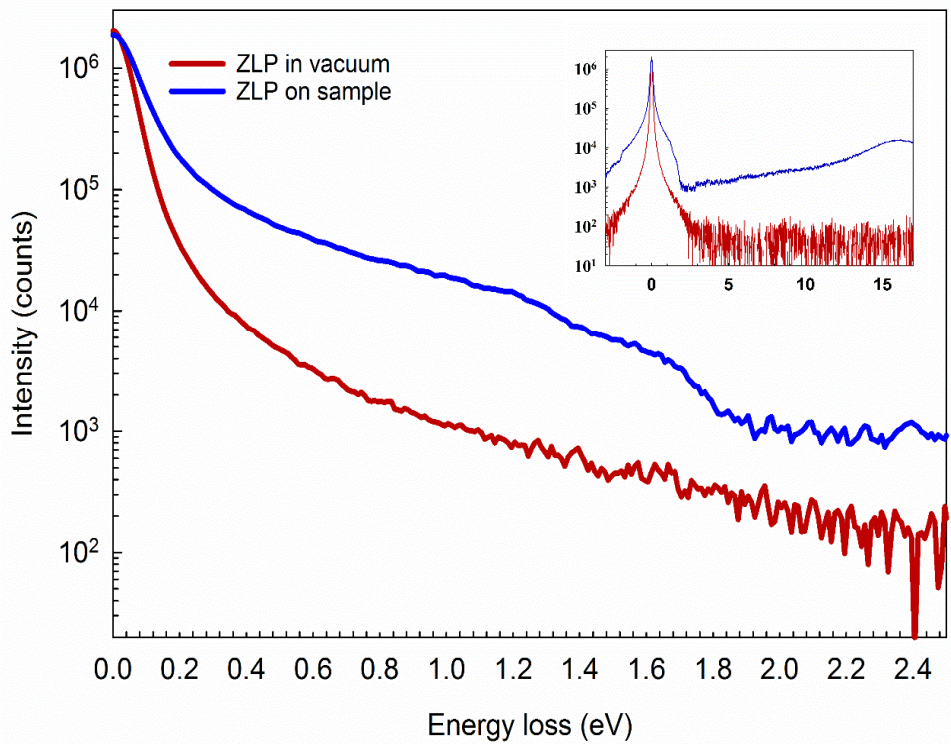

**Fig. 5.** An EEL spectrum acquired in vacuum is compared with the spectrum acquired under the same conditions on the GaAs-based structure. The tail of the ZLP in the loss region before the bandgap energy changes significantly, and the full width at half maximum (FWHM) of the ZLP changes from 0.10 eV in vacuum to 0.14 eV on the sample. A large semi-collection angle of ca. 64.7 mrad was used during the acquisition of the spectra. The whole EELS spectra are shown in the inset. All spectra are shown in logarithmic scale.

### 3.2. Low-Mag on-axis

In Low-Mag STEM mode, the semi-convergence and –collection angles are both very small, as illustrated in Figs. 2 and 3. The region of reciprocal space collected by the spectrometer is now confined to a narrow circle with a radius of 0.05 mrad only. For GaAs, as well as most other practical samples, this region is far from the first BZ boundaries. GaAs oriented close to the [110] zone axis has its X and L points at the BZ boundary along the [002] and [111] directions located 7.39 and 6.40 mrad away from the BZ centre, respectively. Hence, only the centre of the BZ goes onto the spectrometer. This means that any inelastic transitions with a significant momentum transfer, such as the major part of all phonons and intraband excitations and a significant part of all surface and interface plasmons, do not contribute to the EEL spectrum. An on-axis EEL spectrum from GaAs, acquired at 80 kV and with semi-convergence and –collection angles of 0.12 and 0.05 mrad respectively, is shown in Fig. 6. Despite the suppression of a lot of the signal that masks and buries the signal from bandgap excitations, no sign of any optical onset from a direct bandgap at 1.42 eV can be observed directly in the unprocessed spectrum. Attempts to remove the background end up similar to the case where larger semi-angles are used and give a background subtracted signal with an onset energy anywhere between 1 and 2.5 eV, depending on the background subtracted. Obviously, relativistic losses and guided light modes still dominate the intensity in the energy loss region of the bandgap.

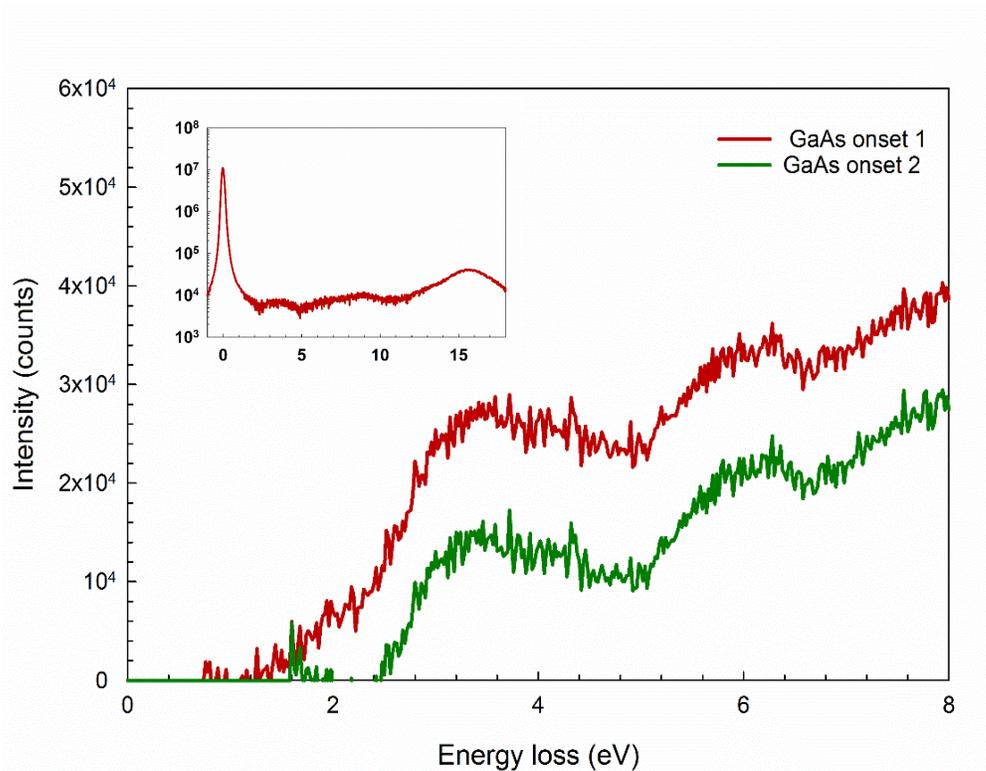

**Fig. 6.** Different bandgap onsets for GaAs in the Low-Mag STEM set-up after background subtraction. Low-Mag on-axis EEL spectrum from GaAs, acquired at 80 kV and with semi-convergence and –collection angles of 0.12 and 0.05 mrad, respectively. No sign of any optical onset from a direct bandgap at 1.42 eV can be observed directly in the unprocessed spectrum. Attempts to remove the background end up similar to the case where larger semi-angles are used and give a background subtracted signal with an onset energy anywhere between 1 and 2.5 eV, depending on the background subtracted. The whole EELS spectrum in logarithm scale is shown in the inset.

### 3.3. Low-Mag off-axis:

The semi-convergence and collection angles in Low-Mag STEM mode are almost two orders of magnitude less than the angle between the BZ centre and the first BZ boundary. The EEL spectra shown in Fig. 7 are acquired by using the projector lenses to shift the diffraction plane and move the entire (000) diffraction disc outside the spectrometer. This shift is just larger than 0.25 mrad to overcome the sum of the experimental semi-convergence and –collection angles of 0.20 and 0.05 mrad respectively (see Fig. 3 b)). Firstly, this off-axis shift is so small that the spectrum is collected from a region very close to the centre of the BZ, and hence contains excitations with very small momentum transfer and that are close to direct bandgap excitations. Secondly, this off-axis shift is still large enough to avoid the forward scattered relativistic losses and guided light modes to contribute to the spectrum. Unprocessed spectra from GaAs, $Al_{0.25}Ga_{0.75}As$ and $Al_{0.25}Ga_{0.75}As_{1-x}N_x$ ($x < 1$) are shown in Fig. 7. The onsets of bandgap excitations are defined by the intersection of a horizontal line describing the background in front of the bandgap and a linear line describing the EEL spectrum in between the bandgap and 2.5 eV (after which the spectrum changes its shape, as expected). The respective bandgaps are now directly read out as 1.42 eV ± 0.02 (GaAs), 1.61 eV ± 0.03 eV ($Al_{0.25}Ga_{0.75}As_{1-x}N_x$) and 1.7 eV ± 0.03 ($Al_{0.25}Ga_{0.75}As$). These values are (close to) equal to the known (GaAs and $Al_{0.25}Ga_{0.75}As$) or expected ($Al_{0.25}Ga_{0.75}As_{1-x}N_x$) direct bandgaps for these materials [17-19].

The off-axis shift of the CBED pattern giving dark field EEL conditions has several implications: The ZLP peak is strongly suppressed, and in our experimental spectra, the maximum intensity of the ZLP has the same order of magnitude as the maximum intensity of the first bulk plasmon peak at about 15 eV (see inset in Fig.7). Whether or not we are on or off one or several of the major Kikuchi bands would strongly determine how much the ZLP is supressed. The spectra in Fig. 7 were acquired without any knowledge of the position of the Kikuchi bands compared to the region in reciprocal space that was probed. However, this is likely to be an important experimental parameter if it is very important to suppress the ZLP as much as possible, for instance if bandgaps in the region below 1 eV are to be determined.
Off-axis conditions combined with very small convergence and collection angles further cause the overall EEL signal to decrease by orders of magnitude compared to on-axis conditions in normal STEM mode where semi-convergence and –collection angles that are tens of mrad. To test the robustness of the measurements above, the monochromator slit was therefore changed from 1 to 2 μm in diameter and the condenser aperture was changed from 30 to 50 μm in diameter to increase the probe current from 50 to 400 pA. The change to a larger slit means that we decrease the EEL energy resolution, and the change of condenser aperture size causes the semi-convergence angle to increase from 0.12 to 0.20 mrad. The spatial resolution in the STEM image however, is not reduced rather improved by the change to a larger condenser aperture since the spatial resolution is no longer limited by the lens aberrations in Low-Mag mode. In addition, the acquisition time was set to a few seconds for each pixel in the lines scans and maps to have sufficient signal to noise. The larger monochromator slit size combined with acquisition times of several seconds increased the FWHM to 0.2 - 0.3 eV. However and as can be seen in Fig. 7, the tail of the ZLP has totally decayed below 1 eV, and the unprocessed spectra have a flat signal for several hundred meV in front of the onset of the bandgap for GaAs.

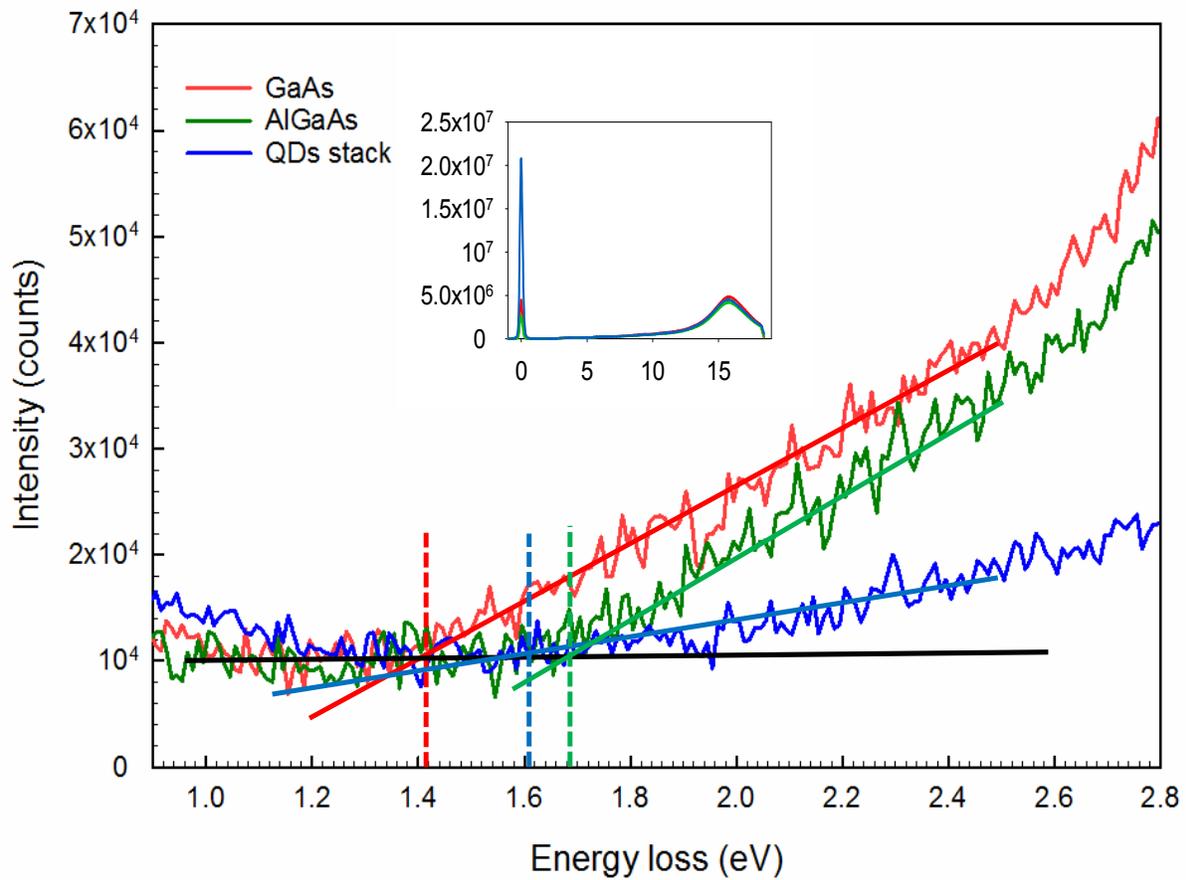

**Fig. 7.** Unprocessed spectra from GaAs, $Al_{0.25}Ga_{0.75}As$ and $Al_{0.25}Ga_{0.75}As_{1-x}N_x$ ($x < 1$) in the Low-Mag off-axis set-up. The EEL spectra are acquired by using the projector lenses to shift the diffraction plane and move the entire (000) diffraction disc outside the spectrometer. The onset of bandgap excitations are defined by the intersection of a horizontal line describing the background in front of the bandgap and a linear line describing the EEL spectrum in between the bandgap and 2.5 eV (after which the spectrum changes its shape, as expected). The respective bandgaps are now directly read out as 1.42 eV ± 0.02 (GaAs), 1.61 eV ± 0.03 eV ($Al_{0.25}Ga_{0.75}As_{1-x}N_x$) and 1.7 eV ± 0.03 ($Al_{0.25}Ga_{0.75}As$). The whole EELS spectra are shown in the inset.

### 3.4. Aloof:

The long-range Coulomb interactions between the fast electrons in the electron beam and the sample electrons that define a scattering centre give a large delocalisation effect, which is utilized in the aloof technique. An electron beam that is positioned several nm away from the sample can still excite inelastic transitions in the sample due to this delocalization. It has been shown that for instance bandgaps and surface plasmons can be excited and detected several tens of nm away from the sample [16]. Furthermore, when the beam never goes through the sample one would expect to suppress the relativistic losses and guided light modes in the spectra. Line scans both parallel and perpendicular to the sample were acquired to try to retrieve the onset of the bandgap in the nearby sample material. However, and independent of the distance from the sample, we were not able to observe any bandgap signal in the spectra. An on-axis spectrum collected 30 nm away from the broken edge of GaAs and taken with semi-convergence and –collection angles of 12.6 and 7.7 mrad at 80 kV is shown in Fig. 8. Even if the tail of the ZLP is strongly suppressed compared to a similar spectrum acquired on the sample, no bandgap signal could be retrieved from the spectrum, so that any kind of screening effect in GaAs-based materials could be the reason.

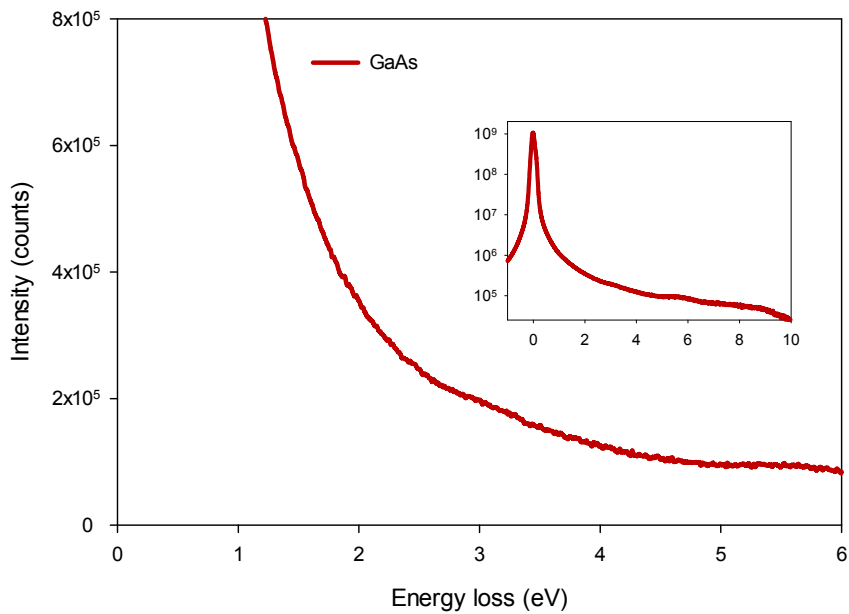

**Fig. 8.** The on-axis spectrum in the Aloof set-up collected 30 nm away from the broken edge of GaAs and taken with semi-convergence and –collection angles of 12.6 and 7.7 mrad at 80 kV. Even if the tail of the ZLP is strongly suppressed compared to a similar spectrum acquired on the sample, no bandgap signal could be retrieved from the spectrum. The whole EELS spectrum in logarithm scale is shown in the inset.

# 4. Conclusions

In this work a range of experimental STEM-EELS set-ups have been systematically explored in order to determine the onset of the optical, direct bandgap of high refractive index GaAs-based materials. Acceleration voltages in the range 60 – 200 kV were combined with a broad range of various semi-convergence and -collection angles, on- and off-axis conditions, aloof scanning, and TEM specimen thicknesses ranging from 30 to 200 nm.

In normal STEM mode and independent of on- or off-axis conditions, specimen thickness and acceleration voltage, the EEL signal in the loss region of the bandgap was totally dominated by contributions from relativistic losses, phonons and excitons, surface losses and guided light modes. The combined signal from these unwanted losses masked the relatively weak signal from bandgap excitations. Furthermore, no background procedure was able to deconvolute the buried bandgap signal from the complex background.

However, we exploited that all relativistic losses and guided light modes are confined inside a narrow forward-scattered solid angle, extending out to a few tens of µrad only. In Low-Mag STEM the semi-convergence and –collection angles can both go far down into the µrad range. Such low semi-angles were combined with off-axis conditions to collect electrons scattered to outside the angular range of relativistic losses and guided light modes. Off-axis conditions further suppressed the ZLP and the tail of the ZLP, which allowed the bandgap to be directly seen and extracted from unprocessed spectra. Off-axis conditions mean that we collect electrons that have received a momentum transfer. However, since the off-axis scattering angles that contribute to the EEL signal are small compared to the distance to any Brillouin zone boundary, the detected bandgap excitations are also expected to be equal or close to equal to the material's direct bandgap. These assumptions were confirmed by measured bandgaps that were all (close to) identical to the known or expected direct bandgaps of the GaAs-based materials.

**Acknowledgements**

The authors would like to thank Professor Shu Min Wang and Mahdad Sadeghi at the Nanofabrication Laboratory at Chalmers University, Sweden for providing the samples. The Norwegian Research Council is acknowledged for funding the HighQ-IB project under contract no. 10415201. The research leading to these results has received funding from the European Union Seventh Framework Programme under Grant Agreement 312483 - ESTEEM2 (Integrated Infrastructure Initiative–I3) through the system of transnational access. R.E. and J.V. acknowledge funding from GOA project "Solarpaint" of the University of Antwerp.